\documentclass[prl, aps, twocolumn, showpacs, showkeys, superscriptaddress]{revtex4}
\usepackage{epsfig}
\usepackage{amssymb}

\begin{document} 
  
\title{Enhancing piezoelectricity through polarization-strain coupling in ferroelectric superlattices} 

\author{Valentino R. Cooper}
\email{coopervr@ornl.gov} 
\affiliation{Materials Science and Technology Division, Oak Ridge National Laboratory\\ Oak Ridge, Tennessee 37831-6114, USA}

\author{Karin M. Rabe} 
\affiliation{Department of Physics and Astronomy, Rutgers University\\
136 Frelinghuysen Rd, Piscataway, New Jersey 08854-8019, USA}

\date{\today}

\begin{abstract}
Short period ferroelectric/ferroelectric BaTiO$_{\rm 3}$
(BTO)/PbTiO$_{\rm 3}$ (PTO) superlattices are studied using density
functional theory.  Contrary to the trends in
paraelectric/ferroelectric superlattices the polarization remains
nearly constant for PTO concentrations below 50\%. In addition, a
significant decrease in the $c/a$ ratio below the PTO values were
observed.  Using a superlattice effective Hamiltonian we predict an
enhancement in the $d_{33}$ piezoelectric coefficient peaking at
$\sim$75\% PTO concentration due to the different polarization-strain
coupling in PTO and BTO layers.  Further analysis reveals that these
trends are bulk properties which are a consequence of the reduced $P$
brought about by the polarization saturation in the BTO layers.
\end{abstract}



\maketitle 

Perovskite superlattices present a new paradigm for engineering
ferroelectrics and piezoelectrics for modern device applications.
Epitaxial strains, resulting from lattice mismatches, and induced
changes in polarization, resulting from electrostatic considerations,
can have significant effects on the macroscopic properties of these
artificial structures.\cite{Dawber05p1083,Lee05p395}
Superlattices combining a paraelectric (PE), such as SrTiO$_{\rm 3}$
(STO), with a ferroelectric (FE), such as PbTiO$_{\rm 3}$ (PTO) or
BaTiO$_{\rm 3}$ (BTO) have been well studied both by experiment and
theory.\cite{Neaton03p1586, Johnston05p100103, Dawber05p177601,
Cooper07p020103R, Bousquet08p732} Their main features can be modeled
by considering them as layers of strained bulk-like material with
appropriate electrostatic boundary conditions \cite{Neaton03p1586}; in
some systems, interface effects have also been shown to be important
in the limit of ultrathin layers\cite{Cooper07p020103R}.  In this
paper, we extend this first-principles modeling approach to
superlattices combining two ferroelectrics, which we expect to exhibit
new features arising from the interplay of the different bulk
spontaneous polarizations and dielectric and piezoelectric
coefficients of the two constituents.

\begin{figure}
\includegraphics[width=\columnwidth]{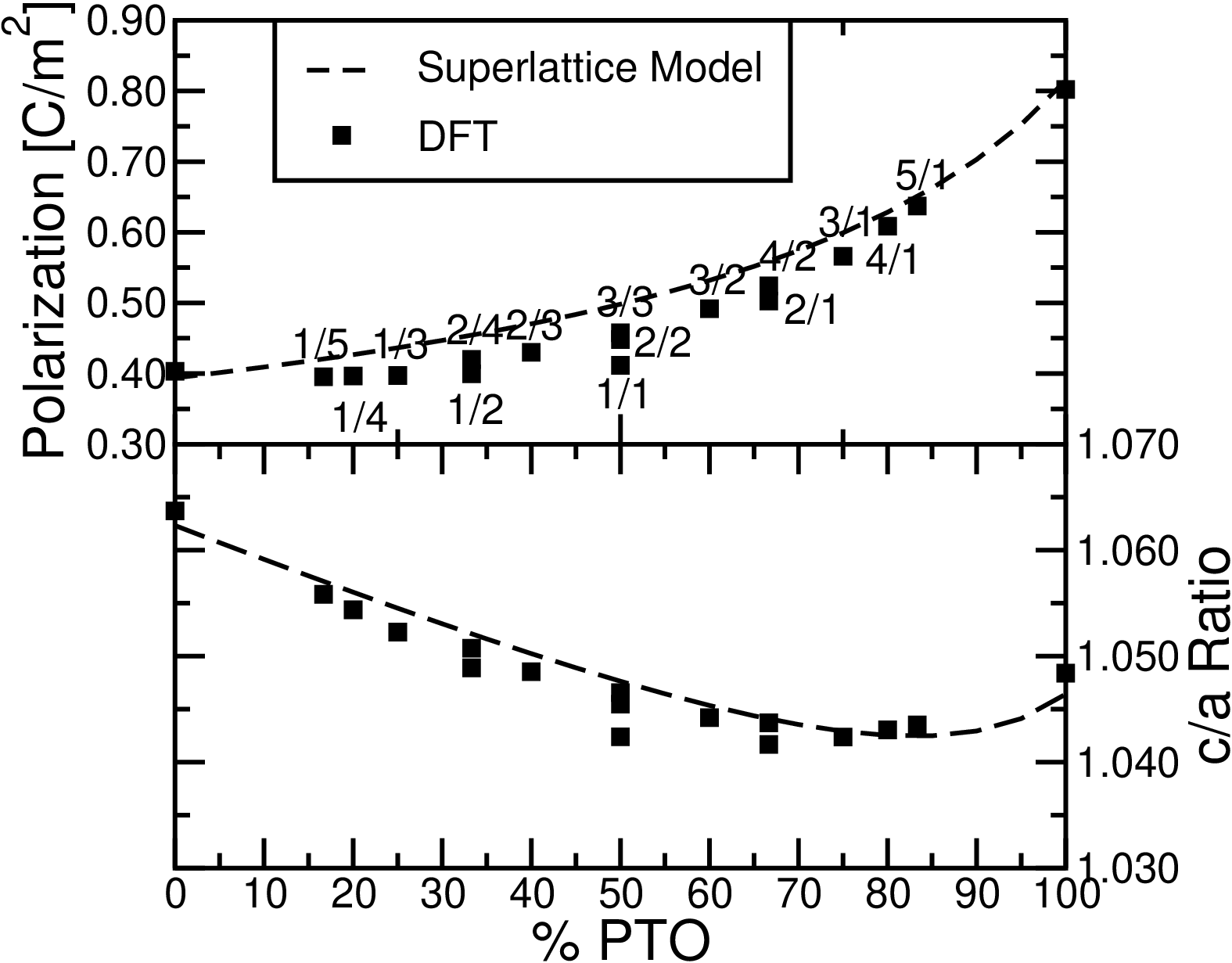}
\caption{\label{PTO_BTOmain} Polarization along the $z$ axis (top) and
  $c/a$ ratio (bottom) as a function of \% PTO for BTO/PTO superlattices
  with varying superlattice periods.  Solid squares represent DFT
  calculations and the dashed lines are the values obtained using the
  superlattice model (see Eqs.~\ref{EffEnth_pure}~
  and~\ref{G_total}).  Numerical model parameters can be found in
  Table~\ref{params}.}
\end{figure}

We perform first-principles calculations, using density functional
theory (DFT)~\cite{Hohenberg64pB864, Kohn65pA1133}, for PTO/BTO
superlattices with m=1-3 and n=1-3 layers of PTO and BTO,
respectively. All calculations used projector augmented wave (PAW)
potentials~\cite{Blochl90p5414, Kresse99p1758} with the Vienna Ab
initio Simulation Package (VASP v4.6.26))~\cite{Kresse96p11169}, with
the local density approximation for the exchange correlation
functional. A 700 eV (22 Ha) cutoff and a 6$\times$6$\times$$l$
k-point mesh were used (where $l$=4 for periods 2 and 3 superlattices,
2 for periods 4 and 5 and 1 for period 6).  For BTO in the tetragonal
$P4mm$ 5-atom unit cell structure, the computed lattice constants are
a=3.947~\AA\ and c=3.996~\AA.  The tetragonal PTO lattice constants
were computed as a=3.867~\AA\ and c=4.033~\AA\ [experiment:
a=3.904~\AA\ and c=4.152~\AA].\cite{Hellwege1981} This agreement is
typical of LDA calculations for ferroelectric perovskites.  In all
superlattice calculations the in-plane lattice constant was
constrained to that of the theoretical value for an STO substrate
(3.863~\AA) while the $c$ lattice vectors were optimized within the
$P4mm$ space group with 1x1 in-plane periodicity.  Additional
calculations for a doubled $\surd$2$\times\surd$2 in-plane unit cell
for m=n=1 show no instability to octahedral tilts of the type
discussed in Ref.~\onlinecite{Bousquet08p732}.  All ionic coordinates
were fully relaxed until the Hellman-Feynman forces on the ions were
less than 5 meV/\AA.  Polarizations were computed using the Berry
phase method~\cite{King-Smith93p1651}.  For bulk PTO, constrained to
the hypothetical STO substrate in-plane lattice constant of 3.863~\AA,
the out-of-plane lattice constant was 4.039~\AA\ and the polarization
was 0.79~C/m$^2$.  The epitaxial constraints placed on the BTO layers
result in a BTO $c/a$ ratio (1.063) much larger than that of the more
polar PTO layers. The BTO polarization was 0.39~C/m$^2$, substantially
enhanced over the computed bulk value of 0.26~C/m$^2$.

Figure~\ref{PTO_BTOmain} shows the dependence of computed polarization
and $c/a$ lattice constant on the concentration of PTO and the
superlattice period of short period PTO/BTO superlattices for the full
set of structures with period $\leq$ 6.  Consistent with the fact that
PTO has a greater polarization than BTO, higher concentrations of PTO
give rise to larger $P$ of the superlattice.  Similarly, decreasing
the content of the larger $c/a$ ratio strained BTO reduces the average
$c/a$ ratio of the superlattice. However, closer examination shows some
unexpected features.  First, the polarization of the superlattices
remains almost constant for PTO concentrations below 50~\% while the
$c/a$ ratio is below, rather than above, that of bulk PTO for a wide
range of PTO concentrations ($\sim$40~\% to 100\%).  In fact, the
derivative of $c/a$ with respect to PTO concentration is positive near
100\%, rather than negative as would be expected from a simple linear
interpolation.

To interpret our DFT data and to distinguish between interface effects
and bulk properties, we constructed a parameterized energy expression
for the superlattice following Refs. ~\onlinecite{King-Smith94p5828}
and ~\onlinecite{Dieguez05p144101}.  The expression is a Taylor
expansion around the cubic perovskite structure in terms of the six
independent components $\eta_i$ of the strain tensor ($i$ is a Voigt
index, $i$ =1--6) and the three Cartesian soft mode amplitude
components $u_{\alpha}$($\alpha=x, y, z$) parameterized from DFT
calculations.  Restricting to $P4mm$ symmetry, the effective
stress-strain elastic enthalpy for a pure component ferroelectric
under epitaxial strain (i.e. $\eta_1$ = $\eta_2$ = $\bar{\eta}$) with
out-of-plane stress, $\sigma_3$, can be written as:

\begin{eqnarray}\label{EffEnth_pure}
\nonumber G(\bar{\eta}, \eta_3, u_z, \sigma_3) = \frac{1}{2}
B_{11} ( 2 \bar{\eta}^2 + \eta_{3}^2) + B_{12} (\bar{\eta}^2 + 2
\eta_{3} \bar{\eta}) +\\
\kappa u_{z}^2 +\alpha u_{z}^4 + 
 \frac{1}{2} 
B_{1xx} \eta_3 u_{z}^2 +B_{1yy} \bar{\eta} u_{z}^2 - \sigma_3\eta_3\;,
\end{eqnarray}

\noindent where $B_{11}$ and $B_{12}$ are related to the elastic
constants of the crystal, $\kappa$ and $\alpha$ are two independent
symmetry-allowed fourth-order coefficients describing the cubic
anisotropy and $B_{1xx}$ and $B_{1yy}$ are the phonon-strain coupling
coefficients.

\begin{figure}[]
\includegraphics[width=\columnwidth]{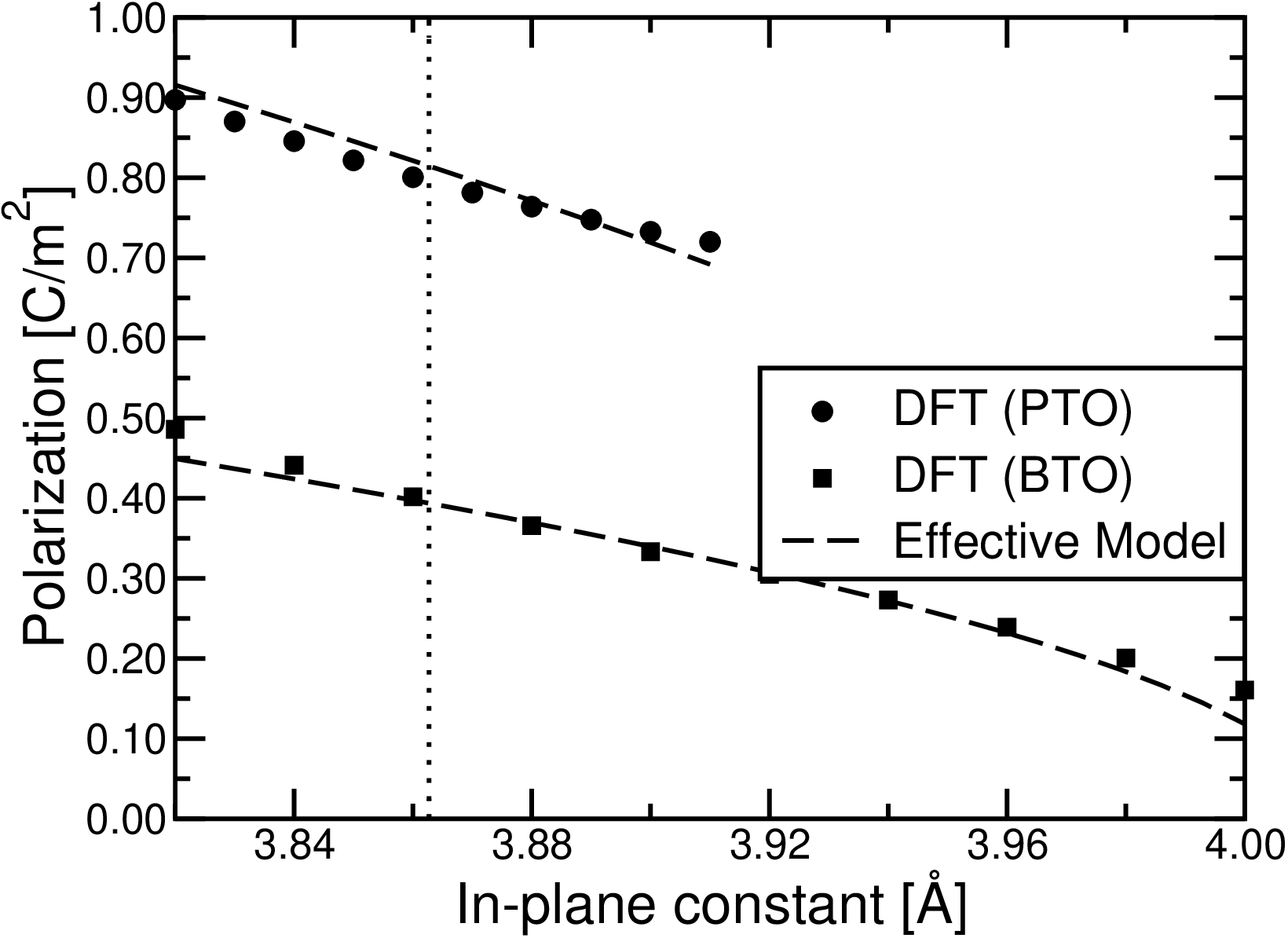}
\caption{\label{BTO_inplane} Polarization along the $z$ axis as a
function of in-plane strain for bulk BTO and PTO.  Solid circles and
squares represent DFT calculations for PTO and BTO,
respectively. Dashed lines are the corresponding values obtained
using Eq. \ref{EffEnth_pure} with the parameters of Table~\ref{params}.}
\end{figure}

Figure \ref{BTO_inplane} shows the polarization versus in-plane
lattice constant for pure component BTO and PTO.  The model parameters
(Table~\ref{params}) were fit to DFT data for PTO with in-plane
lattice constants within $\pm$1$\%$ of the STO in-plane lattice
constant 3.863\AA~. For BTO, the fitting range was extended to include
the computed bulk tetragonal lattice constant of 3.947\AA~.  The
excellent agreement between the model and our DFT calculations is an
indication of the quality of the model.

\begin{table}[b!]
\caption{\label{params} Energy expansion coefficients,
Eq.~\ref{EffEnth_pure}, and soft mode Born effective charges, for PTO
and BTO, in atomic units fit to the DFT results in Fig. \ref{BTO_inplane},
as described in the text.}
\begin{tabular}{ccccccccc}
\hline
\hline
& $B_{11}$ & $B_{12}$ & $B_{1xx}$ & $B_{1yy}$ & $\kappa$ & $\alpha$ &
$Z^*$ & $a_{\rm o}$\\
\hline
BaTiO$_3$ & 5.13 & 3.05 & -1.00 & -0.100 & -0.007 & 0.14 & 9.94 & 7.48 \\
PbTiO$_3$ & 4.69 & 1.15 & -0.705 & 0.207 & -0.0132 & 0.0364 & 9.40 & 7.385 \\
\hline
\hline
\end{tabular}
\end{table}

\noindent For a given $m_{\rm PTO}$/$n_{\rm BTO}$ superlattice we
assume uniform soft mode amplitude $u_z$ within each constituent
layer.  Further, we impose the constraint that the corresponding
polarizations of the two constituent layers $i$ = BTO, PTO,

\begin{equation}\label{layerP}
P_z^i = \frac{e}{\Omega^i}Z^{*,i}u_z^i,
\end{equation}

\noindent are equal, where $e$ is the absolute value of the electron
 charge, $\Omega^i$ is the layer unit cell volume and $Z^{*,i}$ is the
 Born effective charge of the soft mode, so that polarization is
 uniform throughout the superlattice. Strictly speaking, it is the
 displacement field, $D = P + \varepsilon E$, which remains uniform
 throughout the superlattice.  However, this uniform polarization
 approximation has previously been shown to be valid for short-period
 superlattices~\cite{Nakhmanson05p102906, Cooper07p020103R} and is
 supported in the present case by unit-cell-layer polarization
 profiles computed using bulk Born effective charges (not shown).

The effective elastic enthalpy for the two-component-superlattice 
with $m_{\rm PTO}$ layers and $n_{\rm BTO}$ layers is then obtained as:

\begin{equation}\label{G_total}
G^{\rm Total} = \frac{1}{(m+n)}[mG^{\rm PTO} +
nG^{\rm BTO}].
\end{equation}

The superlattice $P$ and $\eta_3$
are determined by minimizing Eq.~\ref{G_total} with respect to
$\eta_3^i$ and $u_z^i$, imposing the uniform polarization approximation
through the relation
\begin{equation}
\label{Gamma}
u_z^{\rm BTO} = \Gamma u^{\rm PTO}_z, {\rm with }\;
\Gamma=\frac{\Omega^{\rm BTO}  Z^{*, \rm PTO}}{\Omega^{\rm PTO} Z^{*, \rm BTO}}. 
\end{equation}



\begin{figure}
\includegraphics[width=\columnwidth]{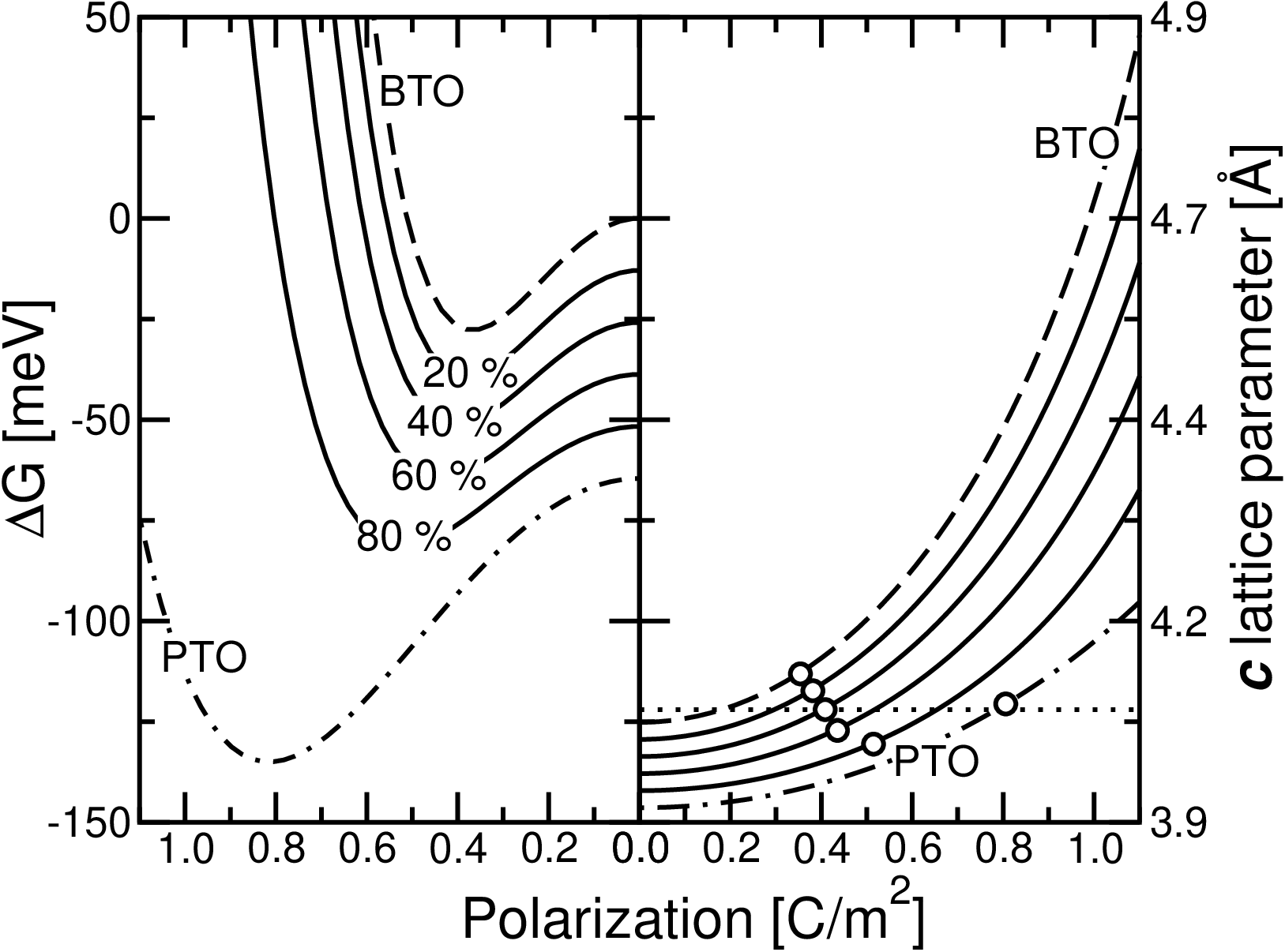}
\caption{\label{fig_d33} Dependence of $d_{33}$ on PTO concentration
  for PTO/BTO superlattices.  The dashed line represent the
  superlattice effective Hamiltonian at bulk STO in plane lattice
  constant of 3.863 \AA\ and the dotted line is are the same for an
  in-plane lattice constant of 3.90 \AA. The solid squares are computed
  from DFT.}
\end{figure}

This model can also be used to compute the piezoelectric coefficient,
$d_{33}$, as:

\begin{eqnarray}\label{d33}
\nonumber\frac{dP}{d\sigma_3} = - \frac{m e Z^{*,PTO} B_{1xx}^{PTO}}{2\Omega^{PTO} u_z^{PTO}(2B_{11}^{PTO}\alpha^{PTO}-\frac{1}{4} (B_{1xx}^{PTO})^2)} \\
- \frac{m e Z^{*,BTO} B_{1xx}^{BTO}}{2\Omega^{BTO} u_z^{BTO}(2B_{11}^{BTO}\alpha^{BTO}-\frac{1}{4} (B_{1xx}^{BTO})^2)}. 
\end{eqnarray}

\noindent Figure~\ref{fig_d33} shows the dependence of $d_{33}$ on
composition as predicted from Eq.~\ref{d33}. The model is in good
agreement with the DFT $d_{33}$ values for bulk BTO (32 pC/N), bulk
PTO (55 pC/N) and the 3 PTO / 1 BTO superlattice (58
pC/N). Surprisingly, we find a 5$\%$ enhancement of the $d_{33}$
coefficient at 75~\% PTO where the $c/a$ is a minimum.

\begin{figure}[b]
\includegraphics[width=\columnwidth]{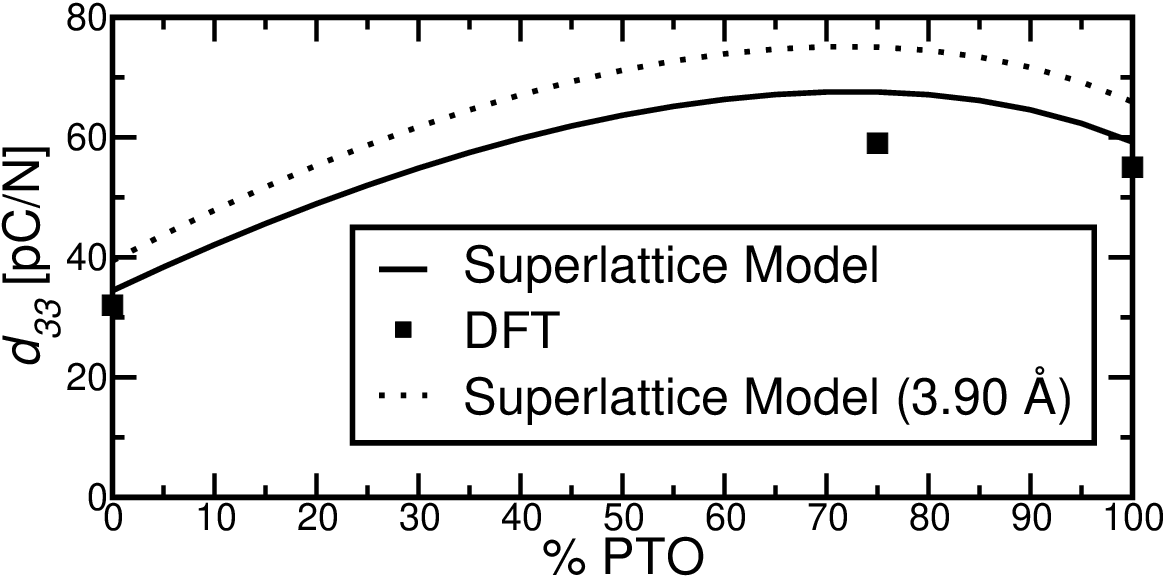}
\caption{\label{PTO_BTO_PEwell} Effective enthalpy (left) and $c$
  lattice parameter (right) as a function of polarization for various
  compositions of PTO/BTO superlattices obtained from the superlattice
  energy functional.  Dashed lines represent pure BTO,
  dashed-dotted lines are for pure PTO and solid lines indicate the
  ferroelectric wells for the superlattices at 20\% PTO
  intervals. Open circles mark the c lattice parameter for each
  composition at the predicted $P$. The dotted line in the right figure
  indicates $c$ for epitaxially-strained tetragonal PTO. Energies are
  relative to epitaxially strained bulk BTO.}
\end{figure}

The good agreement between the DFT results and the superlattice energy functional, shown in Fig.~\ref{PTO_BTOmain}, suggests that the observed
trends in P and $c/a$ are a consequence of bulk electrostatics and strain 
effects, and that interfaces do not play an essential role. 
The success of the model allows us to elucidate the origin of the nearly constant
$P$ for low PTO concentration through an examination of the evolution
of the FE potential energy well (Fig.~\ref{PTO_BTO_PEwell}, left
panel).  A comparison of the FE potential energy wells of the
epitaxially strained parent compounds shows that BTO has a much
stiffer FE well with a minimum at a $P$ of less than half of PTO.  For
polarizations much greater than 0.39 $C/m^2$ the effective enthalpy of
BTO sharply increases, while PTO has a much gentler dependence.  Here
the $P$ in the BTO layers is saturated, requiring larger electric
fields to further polarize them.  Similar polarization saturation
effects have been observed in compressively strained PZT and
PTO~\cite{Lee07p217602}.  In a superlattice, this translates into
the need for a higher percentage of the more polar PTO in order to
increase the macroscopic polarization of the superlattice.  This fact
is borne out in the evolution of the FE wells with \% PTO. Even up to
PTO concentrations of 80\% the FE well still resembles that of BTO,
severely limiting the total $P$ in the superlattice and accounting for
the suppression in expected $P$ enhancement with increasing \%~PTO.

This suppressed layer-by-layer polarization can be linked to the
abnormal decrease of $c/a$ below that of the PTO parent compound
(Fig.~\ref{PTO_BTO_PEwell}) as well as the enhancement in the $d_{33}$
coefficient.  The model indicates that if $P$ were to increase
linearly with increasing PTO concentration then the $c$ lattice
parameter would have a corresponding linear decrease to the PTO
values.  However, as previously stated, the $P$ throughout the
superlattices is severely suppressed due to the presence of the BTO
layers.  Since $P$ is coupled to the $c$ lattice parameter this
results in a drastic reduction in the average $c$ lattice parameter
(see Fig.~\ref{PTO_BTO_PEwell}).  This BTO effect is only overcome at
extremely high concentrations of PTO, where the electric field
generated by the PTO layers are capable of counteracting the BTO
potential energy well constraints.  Furthermore, BTO has a much
steeper dependence of $c$ on $P$ than PTO, i.e. the $c$ lattice
parameter increases more with increasing $P$ in the BTO layers than in
the PTO layers. As such, it is the balance between the stronger
polarization-strain coupling in the BTO layers and the amount of PTO
present to pole these layers which give rise to the peak in $d_{33}$
at $\sim$75\% PTO concentration.  These results suggest that it may be
possible to further enhance $d_{33}$ by reducing the saturation in the
BTO layers either by decreasing the in-plane compressive strain (see
Fig.~\ref{PTO_BTO_PEwell}, dotted line: $a$=3.90\AA) or through other
factors such as intermixing at the interface between PTO and BTO
layers.\cite{Cooper07p020103R}

In conclusion, we have used density functional theory to explore the
dependence of $P$ and $c/a$ as a function of PTO concentration for
short period PTO/BTO superlattices.  Our extensive DFT calculations
reveal two intriguing phenomena: nearly constant $P$ for PTO
concentrations less than 50\% and a dramatic decrease in $c/a$ to
values below that of the smaller PTO.  Using a superlattice effective
Hamiltonian we demonstrate that these trends are a consequence of $P$
saturation in the BTO layers which, by limiting the magnitude of $P$,
results in abnormally large decreases in the $c$ lattice
parameter. The competition between the polarization saturation effects
of BTO and the high polarization in the PTO layers results in a
corresponding peak in the $d_{33}$ coefficients in the PTO/BTO
superlattices, suggesting a new way of enhancing the piezoelectric
properties of the superlattices; other material combinations may
yet be found which show an even greater enhancement based on this
mechanism. In addition, this model can be easily 
extended to multicomponent systems,
different phases and to model strain effects on these phases; allowing
for the easy exploration of materials properties.

We would like to thank David Vanderbilt and Scott Beckman for valuable
discussions. This work was supported by ONR Grant
N0014-00-1-0261. Part of this work was carried out at the Aspen Center
for Physics. Work at ORNL was supported by DOE, Division of Materials
Sciences and Engineering.


\begin{thebibliography}{18}
\expandafter\ifx\csname natexlab\endcsname\relax\def\natexlab#1{#1}\fi
\expandafter\ifx\csname bibnamefont\endcsname\relax
  \def\bibnamefont#1{#1}\fi
\expandafter\ifx\csname bibfnamefont\endcsname\relax
  \def\bibfnamefont#1{#1}\fi
\expandafter\ifx\csname citenamefont\endcsname\relax
  \def\citenamefont#1{#1}\fi
\expandafter\ifx\csname url\endcsname\relax
  \def\url#1{\texttt{#1}}\fi
\expandafter\ifx\csname urlprefix\endcsname\relax\def\urlprefix{URL }\fi
\providecommand{\bibinfo}[2]{#2}
\providecommand{\eprint}[2][]{\url{#2}}

\bibitem[{\citenamefont{Dawber et~al.}(2005{\natexlab{a}})\citenamefont{Dawber,
  Rabe, and Scott}}]{Dawber05p1083}
\bibinfo{author}{\bibfnamefont{M.}~\bibnamefont{Dawber}},
  \bibinfo{author}{\bibfnamefont{K.~M.} \bibnamefont{Rabe}}, \bibnamefont{and}
  \bibinfo{author}{\bibfnamefont{J.~F.} \bibnamefont{Scott}},
  \bibinfo{journal}{Rev. Mod. Phys.} \textbf{\bibinfo{volume}{77}},
  \bibinfo{pages}{1083 } (\bibinfo{year}{2005}{\natexlab{a}}).

\bibitem[{\citenamefont{Lee et~al.}(2005)\citenamefont{Lee, Christen, Chisholm,
  Rouleau, and Lowndes}}]{Lee05p395}
\bibinfo{author}{\bibfnamefont{H.~N.} \bibnamefont{Lee}},
  \bibinfo{author}{\bibfnamefont{H.~M.} \bibnamefont{Christen}},
  \bibinfo{author}{\bibfnamefont{M.~F.} \bibnamefont{Chisholm}},
  \bibinfo{author}{\bibfnamefont{C.~M.} \bibnamefont{Rouleau}},
  \bibnamefont{and} \bibinfo{author}{\bibfnamefont{D.~H.}
  \bibnamefont{Lowndes}}, \bibinfo{journal}{Nature}
  \textbf{\bibinfo{volume}{433}}, \bibinfo{pages}{395} (\bibinfo{year}{2005}).

\bibitem[{\citenamefont{Neaton and Rabe}(2003)}]{Neaton03p1586}
\bibinfo{author}{\bibfnamefont{J.~B.} \bibnamefont{Neaton}} \bibnamefont{and}
  \bibinfo{author}{\bibfnamefont{K.~M.} \bibnamefont{Rabe}},
  \bibinfo{journal}{Appl. Phys. Lett.~} \textbf{\bibinfo{volume}{82}},
  \bibinfo{pages}{1586} (\bibinfo{year}{2003}).

\bibitem[{\citenamefont{Johnston et~al.}(2005)\citenamefont{Johnston, Huang,
  Neaton, and Rabe}}]{Johnston05p100103}
\bibinfo{author}{\bibfnamefont{K.}~\bibnamefont{Johnston}},
  \bibinfo{author}{\bibfnamefont{X.}~\bibnamefont{Huang}},
  \bibinfo{author}{\bibfnamefont{J.~B.} \bibnamefont{Neaton}},
  \bibnamefont{and} \bibinfo{author}{\bibfnamefont{K.~M.} \bibnamefont{Rabe}},
  \bibinfo{journal}{Phys. Rev. B~} \textbf{\bibinfo{volume}{71}},
  \bibinfo{pages}{100103(R)} (\bibinfo{year}{2005}).

\bibitem[{\citenamefont{Dawber et~al.}(2005{\natexlab{b}})\citenamefont{Dawber,
  Lichtensteiger, Cantoni, Veithen, Ghosez, Johnston, Rabe, and
  Triscone}}]{Dawber05p177601}
\bibinfo{author}{\bibfnamefont{M.}~\bibnamefont{Dawber}},
  \bibinfo{author}{\bibfnamefont{C.}~\bibnamefont{Lichtensteiger}},
  \bibinfo{author}{\bibfnamefont{M.}~\bibnamefont{Cantoni}},
  \bibinfo{author}{\bibfnamefont{M.}~\bibnamefont{Veithen}},
  \bibinfo{author}{\bibfnamefont{P.}~\bibnamefont{Ghosez}},
  \bibinfo{author}{\bibfnamefont{K.}~\bibnamefont{Johnston}},
  \bibinfo{author}{\bibfnamefont{K.~M.} \bibnamefont{Rabe}}, \bibnamefont{and}
  \bibinfo{author}{\bibfnamefont{J.~M.} \bibnamefont{Triscone}},
  \bibinfo{journal}{Phys. Rev. Lett.~} \textbf{\bibinfo{volume}{95}},
  \bibinfo{pages}{177601} (\bibinfo{year}{2005}{\natexlab{b}}).

\bibitem[{\citenamefont{Cooper et~al.}(2007)\citenamefont{Cooper, Johnston, and
  Rabe}}]{Cooper07p020103R}
\bibinfo{author}{\bibfnamefont{V.~R.} \bibnamefont{Cooper}},
  \bibinfo{author}{\bibfnamefont{K.}~\bibnamefont{Johnston}}, \bibnamefont{and}
  \bibinfo{author}{\bibfnamefont{K.~M.} \bibnamefont{Rabe}},
  \bibinfo{journal}{Phys. Rev. B~} \textbf{\bibinfo{volume}{76}},
  \bibinfo{pages}{020103(R)} (\bibinfo{year}{2007}).

\bibitem[{\citenamefont{Bousquet et~al.}(2008)\citenamefont{Bousquet, Dawber,
  Stucki, Lichtensteiger, Hermet, Gariglio, Triscone, and
  Ghosez}}]{Bousquet08p732}
\bibinfo{author}{\bibfnamefont{E.}~\bibnamefont{Bousquet}},
  \bibinfo{author}{\bibfnamefont{M.}~\bibnamefont{Dawber}},
  \bibinfo{author}{\bibfnamefont{N.}~\bibnamefont{Stucki}},
  \bibinfo{author}{\bibfnamefont{C.}~\bibnamefont{Lichtensteiger}},
  \bibinfo{author}{\bibfnamefont{P.}~\bibnamefont{Hermet}},
  \bibinfo{author}{\bibfnamefont{S.}~\bibnamefont{Gariglio}},
  \bibinfo{author}{\bibfnamefont{J.-M.} \bibnamefont{Triscone}},
  \bibnamefont{and} \bibinfo{author}{\bibfnamefont{P.}~\bibnamefont{Ghosez}},
  \bibinfo{journal}{Nature} \textbf{\bibinfo{volume}{452}}, \bibinfo{pages}{732
  } (\bibinfo{year}{2008}).

\bibitem[{\citenamefont{Hohenberg and Kohn}(1964)}]{Hohenberg64pB864}
\bibinfo{author}{\bibfnamefont{P.}~\bibnamefont{Hohenberg}} \bibnamefont{and}
  \bibinfo{author}{\bibfnamefont{W.}~\bibnamefont{Kohn}},
  \bibinfo{journal}{Phys. Rev.} \textbf{\bibinfo{volume}{136}},
  \bibinfo{pages}{B864 } (\bibinfo{year}{1964}).

\bibitem[{\citenamefont{Kohn and Sham}(1965)}]{Kohn65pA1133}
\bibinfo{author}{\bibfnamefont{W.}~\bibnamefont{Kohn}} \bibnamefont{and}
  \bibinfo{author}{\bibfnamefont{L.~J.} \bibnamefont{Sham}},
  \bibinfo{journal}{Phys. Rev.} \textbf{\bibinfo{volume}{140}},
  \bibinfo{pages}{A1133} (\bibinfo{year}{1965}).

\bibitem[{\citenamefont{Bl\"{o}chl}(1990)}]{Blochl90p5414}
\bibinfo{author}{\bibfnamefont{P.~E.} \bibnamefont{Bl\"{o}chl}},
  \bibinfo{journal}{Phys. Rev. B} \textbf{\bibinfo{volume}{41}},
  \bibinfo{pages}{5414} (\bibinfo{year}{1990}).

\bibitem[{\citenamefont{Kresse and Joubert}(1999)}]{Kresse99p1758}
\bibinfo{author}{\bibfnamefont{G.}~\bibnamefont{Kresse}} \bibnamefont{and}
  \bibinfo{author}{\bibfnamefont{D.}~\bibnamefont{Joubert}},
  \bibinfo{journal}{Phys. Rev. B} \textbf{\bibinfo{volume}{59}},
  \bibinfo{pages}{1758} (\bibinfo{year}{1999}).

\bibitem[{\citenamefont{Kresse and Furthm\"{u}ller}(1996)}]{Kresse96p11169}
\bibinfo{author}{\bibfnamefont{G.}~\bibnamefont{Kresse}} \bibnamefont{and}
  \bibinfo{author}{\bibfnamefont{J.}~\bibnamefont{Furthm\"{u}ller}},
  \bibinfo{journal}{Phys. Rev. B} \textbf{\bibinfo{volume}{54}},
  \bibinfo{pages}{11169} (\bibinfo{year}{1996}).

\bibitem[{\citenamefont{Hellwege and Hellwege}(1981)}]{Hellwege1981}
\bibinfo{editor}{\bibfnamefont{K.}~\bibnamefont{Hellwege}} \bibnamefont{and}
  \bibinfo{editor}{\bibfnamefont{A.~M.} \bibnamefont{Hellwege}}, eds.,
  \emph{\bibinfo{title}{Landolt-B{\"o}rnstetin}}, vol. \bibinfo{volume}{III}
  (\bibinfo{publisher}{Springer-Verlag}, \bibinfo{address}{Berlin},
  \bibinfo{year}{1981}).

\bibitem[{\citenamefont{King-Smith and Vanderbilt}(1993)}]{King-Smith93p1651}
\bibinfo{author}{\bibfnamefont{R.~D.} \bibnamefont{King-Smith}}
  \bibnamefont{and}
  \bibinfo{author}{\bibfnamefont{D.}~\bibnamefont{Vanderbilt}},
  \bibinfo{journal}{Phys. Rev. B} \textbf{\bibinfo{volume}{47}},
  \bibinfo{pages}{1651} (\bibinfo{year}{1993}).

\bibitem[{\citenamefont{King-Smith and Vanderbilt}(1994)}]{King-Smith94p5828}
\bibinfo{author}{\bibfnamefont{R.~D.} \bibnamefont{King-Smith}}
  \bibnamefont{and}
  \bibinfo{author}{\bibfnamefont{D.}~\bibnamefont{Vanderbilt}},
  \bibinfo{journal}{Phys. Rev. B} \textbf{\bibinfo{volume}{49}},
  \bibinfo{pages}{5828} (\bibinfo{year}{1994}).

\bibitem[{\citenamefont{Di\'{e}guez et~al.}(2005)\citenamefont{Di\'{e}guez,
  Rabe, and Vanderbilt}}]{Dieguez05p144101}
\bibinfo{author}{\bibfnamefont{O.}~\bibnamefont{Di\'{e}guez}},
  \bibinfo{author}{\bibfnamefont{K.~M.} \bibnamefont{Rabe}}, \bibnamefont{and}
  \bibinfo{author}{\bibfnamefont{D.}~\bibnamefont{Vanderbilt}},
  \bibinfo{journal}{Phys. Rev. B} \textbf{\bibinfo{volume}{72}},
  \bibinfo{pages}{144101} (\bibinfo{year}{2005}).

\bibitem[{\citenamefont{Nakhmanson et~al.}(2005)\citenamefont{Nakhmanson, Rabe,
  and Vanderbilt}}]{Nakhmanson05p102906}
\bibinfo{author}{\bibfnamefont{S.~M.} \bibnamefont{Nakhmanson}},
  \bibinfo{author}{\bibfnamefont{K.~M.} \bibnamefont{Rabe}}, \bibnamefont{and}
  \bibinfo{author}{\bibfnamefont{D.}~\bibnamefont{Vanderbilt}},
  \bibinfo{journal}{Appl. Phys. Lett.} \textbf{\bibinfo{volume}{87}},
  \bibinfo{pages}{102906} (\bibinfo{year}{2005}).

\bibitem[{\citenamefont{Lee et~al.}(2007)\citenamefont{Lee, Nakhmanson,
  Chisholm, Christen, Rabe, and Vanderbilt}}]{Lee07p217602}
\bibinfo{author}{\bibfnamefont{H.~N.} \bibnamefont{Lee}},
  \bibinfo{author}{\bibfnamefont{S.~M.} \bibnamefont{Nakhmanson}},
  \bibinfo{author}{\bibfnamefont{M.~F.} \bibnamefont{Chisholm}},
  \bibinfo{author}{\bibfnamefont{H.~M.} \bibnamefont{Christen}},
  \bibinfo{author}{\bibfnamefont{K.~M.} \bibnamefont{Rabe}}, \bibnamefont{and}
  \bibinfo{author}{\bibfnamefont{D.}~\bibnamefont{Vanderbilt}},
  \bibinfo{journal}{Phys. Rev. Lett.} \textbf{\bibinfo{volume}{98}},
  \bibinfo{pages}{217602} (\bibinfo{year}{2007}).

\end{thebibliography}
\end{document}